\documentclass[10pt,letterpaper]{article}
\usepackage[top=0.85in,left=2.75in,footskip=0.75in,marginparwidth=2in]{geometry}

\usepackage[utf8]{inputenc}

\usepackage{cite}

\usepackage{nameref,hyperref}

\usepackage[right]{lineno}

\usepackage{microtype}
\DisableLigatures[f]{encoding = *, family = * }

\raggedright
\usepackage{changepage}

\usepackage[aboveskip=1pt,labelfont=bf,labelsep=period,singlelinecheck=off]{caption}

\makeatletter
\renewcommand{\@biblabel}[1]{\quad#1.}
\makeatother

\usepackage{lastpage,fancyhdr,graphicx}
\usepackage{epstopdf}
\fancyheadoffset[L]{2.25in}
\fancyfootoffset[L]{2.25in}

\usepackage{color}

\definecolor{Gray}{gray}{.25}

\usepackage{graphicx}

\usepackage{sidecap}

\usepackage{wrapfig}
\usepackage[pscoord]{eso-pic}
\usepackage[fulladjust]{marginnote}
\reversemarginpar

\begin{document}
\vspace*{0.35in}

\begin{flushleft}
{\Large
\textbf\newline{On the structures and stabilities of B$_7$Cr$_2$ clusters: A DFT study}
}
\newline
\\
P. L. Rodr\'iguez-Kessler\textsuperscript{1,*}
\\
\bigskip
\bf{1} Centro de Investigaciones en \'Optica A.C., Loma del Bosque 115, Lomas del Campestre , Leon, 37150, Guanajuato, Mexico.
\bigskip
*plkessler@cio.mx

\end{flushleft}

\section*{Abstract}
 In this work, we employ density functional theory (DFT) to explore the structure of boron clusters doped with two chromium atoms (B$_7$Cr$_2$). The results show that the most stable structure is a bipyramidal configuration formed by a B$_7$ ring coordinated with two metallic atoms, similar to recently reported B$_7$M$_2$ analogues. The structural and electronic properties reveal remarkable differences between the global minimum and the higher energy isomers.

\section*{Introduction}

In recent decades, a significant number of theoretical and experimental studies have been devoted to understanding the structure-performance properties of charged and neutral clusters.\cite{ZHOU2006448,TARRAT2017102,Zuo_2024,Chen_2023,10.3389/fchem.2022.841964,molecules29143374} Boron clusters, in particular, are widely studied in the field of materials science and nanotechnology due to their electron-deficient nature and their ability to form various stable and metastable structures. When these clusters are doped with other elements (i.e., foreign atoms are introduced into the cluster), their electronic, magnetic, and chemical properties can be significantly altered, leading to a wide range of potential applications. Although substantial progress has been made in recent years in the experimental and theoretical studies of transition-metal-doped boron clusters, research on boron clusters doped with two transition metal atoms remains relatively limited.\cite{molecules28124721} An experimental and theoretical work on Ta$_2$B$_6^{-/0}$ clusters in 2014 revealed that these clusters adopt a bipyramidal structure, with a B$_6$ ring sandwiched between two Ta atoms.\cite{https://doi.org/10.1002/anie.201309469} In particular, studies on M-doped boron clusters involving alkaline earth metals (M = Be, Mg, Ca) in a specific M$_{2}$B$_7$ composition, particularly Be$_2$B$_7$, have shown exceptional hydrogen storage density.\cite{LI202325821}. Further, by considering the first row transition metals (TM=Fe, Co, Ni) on M$_2$B$_7$ cluster, similar structural patterns have been identified. Moreover these clusters showed plausible reversible hydrogen storage properties. Given the potential properties of two-metallic boron-doped clusters, this work performs structure searches to investigate the most stable structures of Cr$_2$B$_7$ clusters. The stability, vibrational, and electronic properties are discussed. The data provided in this preprint serves as a basis for further investigation into specific potential applications of these clusters.\cite{HUANG201721086,LI202325821,HUANG201721086,GUO2023106216}

\section*{Materials and Methods}

Calculations performed in this work are carried out by using density functional theory computations (DFT) as implemented in the Orca quantum chemistry package\cite{10.1063/5.0004608}. The Exchange and correlation energies are treated by using the PBE0 functional,{\cite{10.1063/1.472933,10.1063/1.478522} in combination with the Def2TZVP basis set. The atomic positions are relaxed self-consistently by a Quasi-Newton method using the BFGS algorithm. The SCF convergence criteria for geometry optimizations are achieved when the total energy difference is smaller than 10$^{-8}$ au, by setting the keyword TightSCF in the input. The  Van  der  Waals  interactions  are  included in the exchange-correlation PBE0 functional with empirical dispersion corrections of Grimme DFT-D3(BJ). {\cite{https://doi.org/10.1002/jcc.21759}} 


\section*{Results}

The structures of the B$_7$Cr$_2$ clusters are obtained using a modified basin-hoping structure search method over seven generations, as discussed in previous works.\cite{D0CP06179D,OLALDELOPEZ2024} Figure~\ref{figure_struc} shows the most stable structure and representative isomers, which are labeled using the {\bf nM}2{\bf .y} notation, where {\bf n} is the number of boron atoms, M is the Cr dopant, and {\bf y} stands for the isomer number in increasing energy ordering.\cite{C5CP01650A}

\begin{figure*}[h!]
\Large
  \vspace{.5cm}
\includegraphics[scale=0.55]{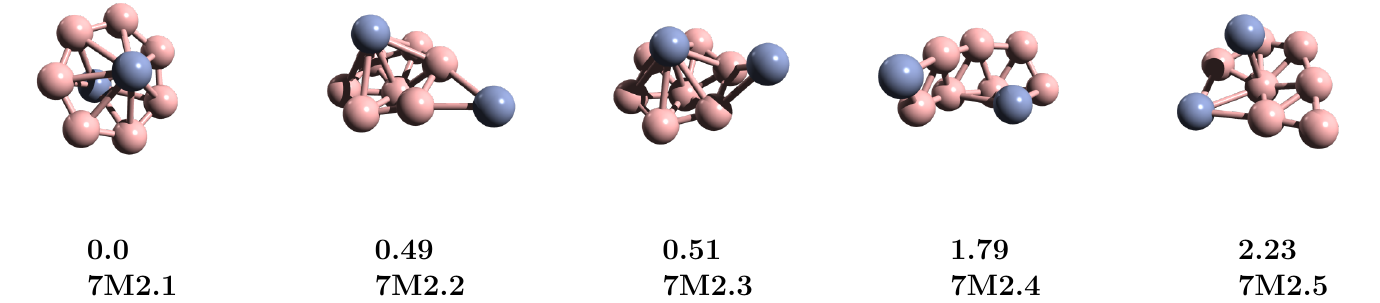}
       \caption{\label{figure_struc}Lowest energy structures for B$_7$Cr$_2$ clusters. For each structure, the relative energy (in eV) and isomer label are given. }
\end{figure*}

The results show that the most stable structure {\bf 7M2.1} (with M=Cr) is an heptagonal bipyramid in the doublet state. The average Cr-B bond distance amounts 2.29 \AA, which is comparable with a previous report on {\bf 7M2} (M=Fe, Co, Ni) clusters, amounting 2.17-2.19 {\AA} for the M-B bonds.\cite{OLALDELOPEZ2024} To characterize the stability of the clusters, Figure~\ref{figure_ECN} shows the isomer distribution as a function of the effective coordination number (ECN) and the average bond distance (d$_{av}$) parameters, which are obtained according to previous reports.\cite{doi:10.1021/acs.jpcc.5b01738} The results show no apparent correlation with ECN, however, the d$_{av}$ indicates that the stability of the most stable cluster is favored at a specific distance of about 1.72 {\AA} while the isomers at higher energies lie in the range of 1.75-1.78 \AA. 

\begin{figure}[h!]
\begin{center}
\scriptsize
 \includegraphics[scale=0.45]{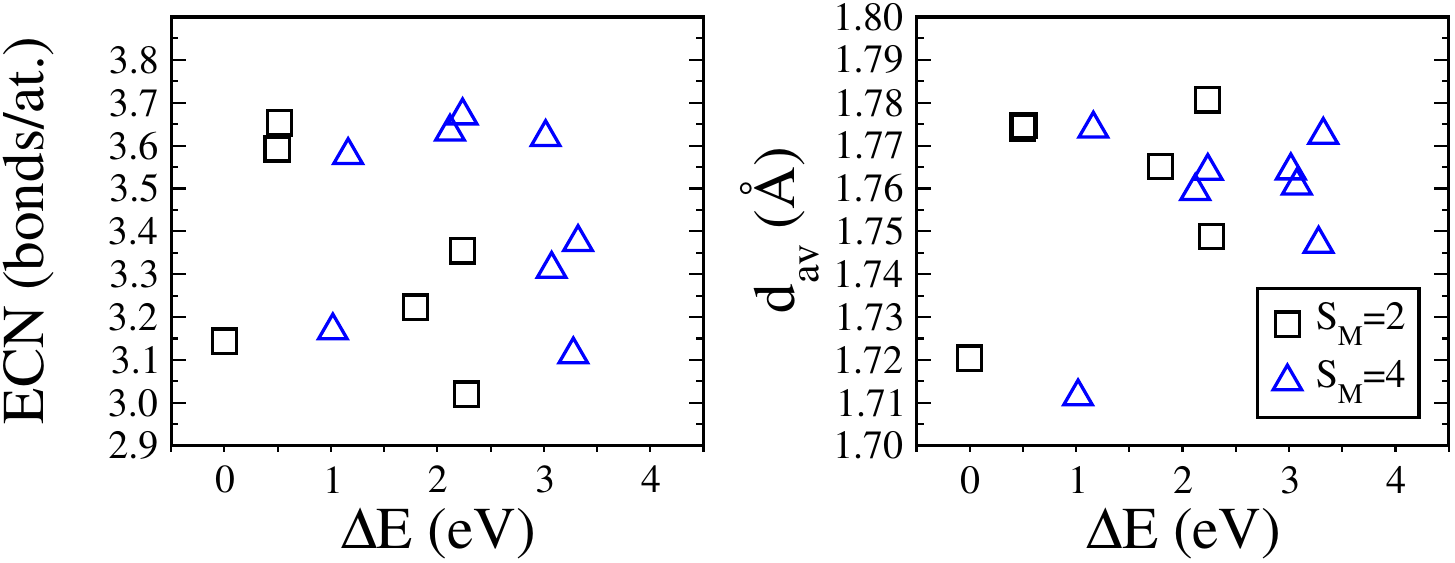}
	\caption{\label{figure_ECN}Effective coordination number and average bond distance as a function of the relative energy ($\Delta{E}$) of B$_7$Cr$_2$ clusters. The spin multiplicities (S$_M$) are denoted by squares and circles, respectively.}
\end{center}
\end{figure}

In order to provide fingerprints for structural identification, we have further calculated the infrared spectra (IR) of the two most stable clusters. The characteristic peaks for {\bf 7Cr2.1} and {\bf 7Cr2.2} clusters were found at 349.35, and 681.88 cm$^{-1}$, suggesting remarkable differences in their structures. The IR spectra of the clusters show significant dispersion for {\bf 7Cr2.2}, while it is more localized for {\bf 7Cr2.1} (Figure~\ref{figure_IR}). The lowest vibrational frequencies are found at 109.18, and 64.20 cm$^{-1}$, while the highest vibrational frequencies are at 1366.26, and 1195.07 cm$^{-1}$, respectively. 

\begin{figure}[h!]
\begin{center}
\scriptsize
 \includegraphics[scale=0.45]{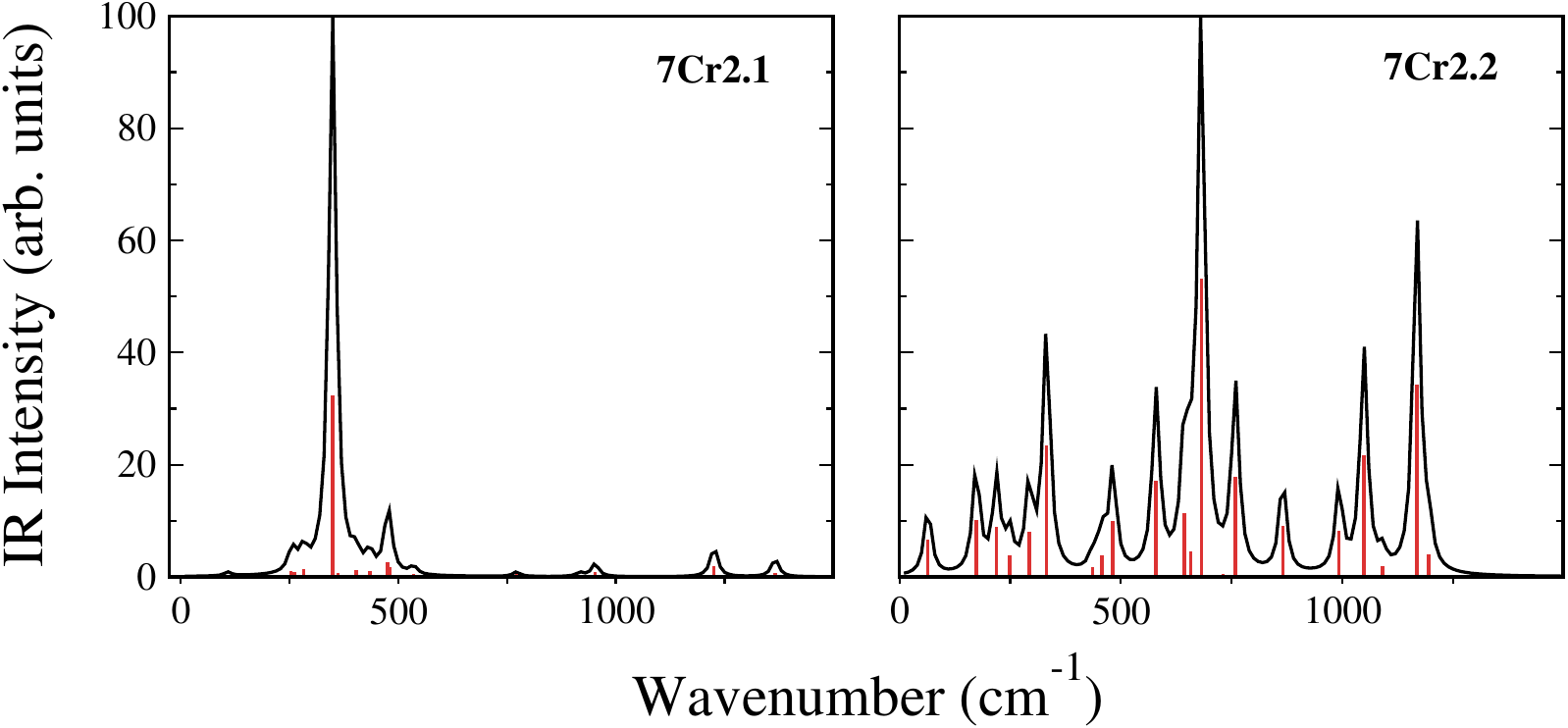}
	\caption{\label{figure_IR}IR spectra for {\bf 7Cr2.1} and {\bf 7Cr2.2} clusters obtained at the PBE0/Def2TZVP level.}
\end{center}
\end{figure}



The ionization energy (IP) and electron affinity (EA) parameters are important physical quantities reflecting the electronic properties of clusters.\cite{SHEN2021125134,LAI2021109757,https://doi.org/10.1002/jcc.27197,DIE2020113805} The calculated values are shown in Table~\ref{table_1}, while their formulas and definitions are available elsewhere.\cite{doi:10.1021/acs.jpcc.9b03637,RODRIGUEZKESSLER2020137721,RODRIGUEZKESSLER201555,RODRIGUEZKESSLER2024122062,doi:10.1021/acs.jpcc.7b05048,D0CP04018E,Kessler2023,D3CP04455F,RODRIGUEZKESSLER2023121620,RODRIGUEZKESSLER201820636,10.1063/1.4935566,doi:10.1021/acs.jpcc.7b05048} The most stable cluster {\bf 7Cr2.1} has lower IP and EA than its isomers, suggesting special reactivity. The structure of the {\bf 7Cr2.1} global minimum is asymmetrical with C$_1$ point group, while higher symmetric configurations were found to be unstable.

	\begin{table}[h!]
	{
 \caption{\label{table_1}{The symmetry point group, electronic state, ionization energy, electron affinity, chemical hardness and chemical potential of {\bf 7Cr2.y} (y=1-5) clusters. The energy is given in eV.}}
\centering
\small
\def\arraystretch{1.1}
\begin{tabular}{p{1.0cm}p{1.0cm}p{1.0cm}p{1.0cm}p{1.0cm}p{1.0cm}p{1.0cm}}
	{cluster} &  Sym  & E$_{state}$	   &  IP     & EA    & $\eta$ & $\mu$    \\ \hline
 {\bf 7Cr2.1} & C$_{1}$ & $^2$A       &  11.80  & 2.83  &  7.31  &  4.48  \\  
 {\bf 7Cr2.2} & C$_s$ & $^2$A$^\prime$ &  12.94  & 4.74  &  8.84  &  4.10  \\
 {\bf 7Cr2.3} & C$_1$ & $^2$A          &  12.37  & 4.02  &  8.20  &  4.17   \\
 {\bf 7Cr2.4} & C$_1$ & $^2$A          &  12.88  & 4.02  &  8.61  &  4.27   \\
 {\bf 7Cr2.5} & C$_1$ & $^2$A          &  12.16  & 4.34  &  8.12  &  4.04   \\
\end{tabular}
	
	}
\end{table}

To further explore the chemical stability of the clusters, we have determined the chemical hardness ($\eta$) and chemical potential ($\mu$), which can be derived from the IP and EA values.\cite{MORATOMARQUEZ2020137677,https://doi.org/10.1002/adts.202100283} The chemical hardness ($\eta$) shows a similar trend compared to IP, which is consistent with the open shell configuration. On the other hand, higher values of chemical hardness ($\eta$) favor the shell closure of clusters.\cite{RODRIGUEZKESSLER2024122062,RODRIGUEZKESSLER2023116538} The chemical potential ($\mu$) is larger than that of the highest energy isomers.

\section*{Conclusions}
In this work, we employed density functional theory (DFT) to explore the structure of boron clusters doped with two metallic atoms (B$_7$Cr$_2$). The results showed that the most stable structure was a bipyramidal configuration, or inverse sandwich structure. Remarkable differences between the global minimum and the higher energy isomers were found. The data provided in this preprint serve as a basis for further investigation into specific potential applications of these clusters. Moreover. further analyses, including the density of states and chemical bonding, will enrich these preliminary results. 


\section*{Supporting Information}


\subsection*{Cartesian coordinates of the B$_7$Cr$_2$ clusters, obtained by using the Orca program and the PBE0/Def2-TZVP approach. For each cluster, the charge state and spin multiplicity are given.}

{\bf \hspace{0.34cm} 7Cr2.1} 0,  2\\
  Cr\hspace{1cm}  -1.50386015016814  \hspace{1cm}   -0.00663717184240 \hspace{1cm}     0.00037526046763\\
  Cr\hspace{1cm}  1.35167915082806\hspace{1cm}      0.00985075549025\hspace{1cm}     -0.00165629475579\\
  B\hspace{1cm}   0.03370833436704\hspace{1cm}     -0.93382475693475\hspace{1cm}     -1.53838086030738\\
  B\hspace{1cm}   -0.01929103416049\hspace{1cm}      0.62268555971966\hspace{1cm}     -1.63873154797093\\
  B \hspace{1cm}  0.00414027617844   \hspace{1cm}  -1.29406165240130  \hspace{1cm}    1.20269871500671\\
  B \hspace{1cm}  0.06832390987082  \hspace{1cm}   -1.81085919401098  \hspace{1cm}   -0.26231281059485\\
  B \hspace{1cm}  0.01725491629219  \hspace{1cm}    1.71803391783632 \hspace{1cm}    -0.52802246490135\\
  B\hspace{1cm}   0.05073555754976  \hspace{1cm}    1.52810081809249  \hspace{1cm}    1.00857527387998\\
  B \hspace{1cm}  -0.00269096075770  \hspace{1cm}    0.16671072405070  \hspace{1cm}    1.75745572917596\\
\vspace{0.5cm} 
{\bf 7Cr2.2} 0,  2\\
 Cr\hspace{1cm}  -1.02840514404939   \hspace{1cm}   0.31469970303970  \hspace{1cm}   -1.26651574488096\\
 Cr\hspace{1cm}  2.60298697563834   \hspace{1cm}  -1.35211962384197  \hspace{1cm}   -1.00657654534875\\
  B\hspace{1cm}   -1.52302636595690 \hspace{1cm}     0.12907938993077  \hspace{1cm}    1.05828635953117\\
  B \hspace{1cm}  0.13564790954583  \hspace{1cm}    0.15110305603492 \hspace{1cm}     0.68982177000907\\
  B\hspace{1cm}   0.47849232940065 \hspace{1cm}     1.60337792120717  \hspace{1cm}   -0.12055168610047\\
  B\hspace{1cm}   0.43904320632971  \hspace{1cm}   -1.13348287778250   \hspace{1cm}  -0.35815271417171\\
  B\hspace{1cm}   -0.88027530578354  \hspace{1cm}   -1.20273889939049  \hspace{1cm}    0.53530000256191\\
  B \hspace{1cm}  -0.85650506104247   \hspace{1cm}   1.52108368440654  \hspace{1cm}    0.72022536513685\\
  B\hspace{1cm}   1.29448944490473  \hspace{1cm}    0.29943848056049  \hspace{1cm}   -0.56994988044374\\

\section*{Acknowledgments}
P.L.R.-K. would like to thank the support of CIMAT Supercomputing Laboratories of Guanajuato and Puerto Interior. 

\nolinenumbers

\bibliography{mendeley}

\bibliographystyle{abbrv}

\end{document}